\documentclass[sigconf,screen]{acmart}

\AtBeginDocument{%
  \providecommand\BibTeX{{%
    \normalfont B\kern-0.5em{\scshape i\kern-0.25em b}\kern-0.8em\TeX}}}

\setcopyright{rightsretained}
\copyrightyear{2023}
\acmYear{2023}
\acmDOI{10.5281/zenodo.8372449}

\acmConference[MuRS '23]{Music Recommender Systems Workshop at the 17th ACM Recommender Systems Conference (RecSys'23)}{September 19, 2023}{Singapore}
\acmPrice{} \acmISBN{}

\begin{document}

\title{Leveraging Negative Signals with Self-Attention for Sequential Music Recommendation}

\author{Pavan Seshadri}
\email{pseshadri9@gatech.edu}
\orcid{0009-0008-7838-9614}
\affiliation{%
  \institution{Georgia Institute of Technology, Music Informatics Group}
  \city{Atlanta}
  \state{Georgia}
  \country{USA}
}
\author{Peter Knees}
\email{peter.knees@tuwien.ac.at}
\orcid{0000-0003-3906-1292}
\affiliation{%
  \institution{TU Wien, Faculty of Informatics}
  \city{Vienna}
  \country{Austria}
}

\renewcommand{\shortauthors}{Seshadri et al.}

\begin{abstract}
Music streaming services heavily rely on their recommendation engines to continuously provide content to their consumers. Sequential recommendation consequently has seen considerable attention in current literature, where state of the art approaches focus on self-attentive models leveraging contextual information such as long and short-term user history and item features; however, most of these studies focus on long-form  content domains (retail, movie, etc.) rather than short-form, such as music. Additionally, many do not explore incorporating negative session-level feedback during training. In this study, we investigate the use of transformer-based self-attentive architectures to learn implicit session-level information for sequential music recommendation. We additionally propose a contrastive-learning task to incorporate negative feedback (e.g skipped tracks) to promote positive hits and penalize negative hits. This task is formulated as a simple loss term that can be incorporated into a variety of deep-learning architectures for sequential recommendation. Our experiments show that this results in consistent performance gains over the baseline architectures ignoring negative user feedback.
\end{abstract}

\ccsdesc[500]{Information systems~Recommender systems}
\ccsdesc[500]{Information systems~Music retrieval}

\keywords{Sequential Recommendation, Music Recommendation, Self-attention, Contrastive Learning}



\maketitle

\section{Introduction}

Recommendation systems have become integral to streaming services such as Spotify, Apple Music, Deezer, etc., and by proxy, the music industry as a whole. As the music streaming business model relies on continual user engagement and activity, consistent music discovery is an essential service. Sequential music recommendation is one such task in this domain, where given a current user session (i.e a current sequence of tracks listened to by a user), a system extends the session by recommending the user the next track. Within the music domain, sequential recommendation is generally split into two categories, \textit{next song recommendation (NSR)}, and \textit{automatic playlist continuation (APC)}. These two tasks can be learned in a similar manner from playlist and listening history information, but they differ in output length: APC aims to extend the session or playlist by an arbitrary length, while NSR only aims to provide the next relevant song in sequence \cite{10.3389/fams.2019.00044}. For this study we focus specifically on NSR.

\indent Music recommendation differs from other well-studied domains of recommendation (retail, movies, games, etc.) in a number of important ways. Singular music tracks generally are short and easily consumed, necessitating a thorough understanding of a user's preferences in order to provide both breadth and depth over a large quantity of relevant recommendations \cite{schedl2015music}. Robust music recommendation systems often leverage previous consumer history to learn user preferences through methods such as collaborative filtering \cite{knees2013survey}; however these approaches fall victim to the cold start problem \cite{schedl2015music}: for new users or new tracks, the recommendation model does not have any usable information and must guess preferences until the user and/or track has interacted with the system enough to learn a profile \cite{https://doi.org/10.48550/arxiv.1708.05031}.
\newline \indent Sequential recommendation in general can alleviate this issue by learning session-level relationships instead of, or in tandem with user-level relationships. By learning session item relationships from sequential interaction, item profiles can be rapidly built as they interact with the system as the recommendation engine can compare user sessions directly rather than using aggregate statistics via collaborative filtering, which takes much more data to build robust representations \cite{https://doi.org/10.48550/arxiv.1708.05031}. 
\newline \indent This study aims to leverage implicit and explicit signals present within listening sessions to learn robust profiles for sequential recommendations. 
Prior work has considered direct incorporation of user feedback for ad-hoc adjustments based on content and context similarity, e.g. \cite{elias_pampalk_2005_1414932,yajie_hu_2011_1418301}.
In this work, we investigate learning session-level information via transformer-based architectures, influenced by SoTA methods for sequential retail recommendation, as well as incorporating user feedback through a learned contrastive task. To our best knowledge, learning from negative signals/user feedback has not been explored thoroughly for sequential music recommendation due to a lack of public data containing thorough user feedback. Many public music recommendation datasets, such as Lastfm-1K \cite{Celma:Springer2010}, were collected before the streaming boom, where logged listening history would primarily be sourced from user creation, leading to a low source of negative signals. For this study, we employ the Music Streaming Sessions Dataset from Spotify \cite{DBLP:conf/www/BrostMJ19}. Since many of the interactions present are from programmatic or expert curation, rather than user curation, they can be considered as exploration events where the user reacts positively (listens to track in entirety) or negatively (skips track). This provides a rich amount of negative samples to learn effective session-level representations from.

\section{Related Work}
Sequential recommendation systems can generally be divided into two types: \textit{session-aware} systems leverage session-level history from identifiable users, while \textit{session-based} systems ignore user-labels and aim to build user-agnostic representations using solely discrete sessions.  \cite{10.1145/3109859.3109896}. In this study, we investigate a \textit{session-based} system that implicitly learns a user profile through anonymous listening sessions. \newline \indent
Several session-based approaches have been proposed for retail recommendation tasks. CASER \cite{https://doi.org/10.48550/arxiv.1809.07426} and NextItNet \cite{10.1145/3289600.3290975} leverage convolutional filters to learn sequential representations. BERT4Rec \cite{https://doi.org/10.48550/arxiv.1904.06690} leverages the bidirectional attention mechanism from BERT \cite{https://doi.org/10.48550/arxiv.1810.04805} to learn a robust vocabulary of items for sequential recommendation. \newline \indent
Several sequential based approaches have been proposed for music recommendation tasks incorporating a variety of information to drive recommendation \cite{10.3389/fams.2019.00044}. Most of such approaches leverage contextual and/or content features, largely by extensive user profiles and music tags. Relevant work for these respective approaches include CoSERNN \cite{10.1145/3383313.3412248}, and Online Learning to Rank for Sequential Music Recommendation \cite{10.1145/3298689.3347019}. The former leverages contextual information such as device used, time of day during recommendation, etc. to drive contextual user-sequential embeddings for sequential recommendation, while the latter leverages content features via music tags for an online learning to rank scheme. For a study closest to our task, Wen et. al investigate leveraging implicit user feedback immediately after click for video and music recommendation, and find performance gains incorporating this information into a variety of recommendation approaches \cite{10.1145/3298689.3347037}. Most state-of-the-art sequential music recommendation approaches leverage several types of information that often are not present in public datasets (e.g lyrics, user contextual/demographic information, music tags, etc.). It would be increasingly difficult to re-implement and test these systems in a cold-start or academic setting due to the amount and variety of data required. Our approach aims to alleviate this data issue by taking advantage of implicit relationships from data present solely in listening sessions of songs, namely item labels and timestamps of user events. We additionally do not take into account long-term user history due to a lack of user labels; Thus, we focus on creating a \textit{session-based} system.

\section{Method}
\subsection{Problem Statement}
In our scenario, we define a session $S$ of length $K$ and set of possible tracks $t \in T$ for user $u$. Track $t_{i}$, where $t_{1, 2, ... K} \in S$ represents the track at each time step $i$ in session $S$, where $i \in [1 \dots K]$. Generally, the task of a sequential recommendation system is to predict the desired next item $h_{i}$ at time step $i + 1$ for each $t_{i} \in S$, given an interaction history $S_{i}$, where  $S_{i} = \{t_{a} \in S \mid a \leq i\}$. \\ 
\indent For negative feedback-agnostic sequential recommendation systems (i.e where the user has not explicitly responded negatively to any item), we define $h_{i}$ for track $t_{i}$ as the next track in the sequence, $t_{i + 1}$.

\indent For our feedback-aware system, we define the set of positive examples (no-skip) as $P$ and negative examples (skipped tracks) as $N$ per sequence $S$, such that:
\begin{center}
$p_{j} \in P$, $n_{k} \in N$, and all $p_{j}, n_{k} \in S$
\end{center}

where $j, k$ correspond to the time step of each example in session $S$. Additionally for clarity, we define $I_{P}$ and $I_{N}$ as the set of time steps for all positive and negative examples, respectively, where $j \in I_{P}$ and $k \in I_{N}$.  For any track $t_{i}$, we define the desired next track $h_{i}$ as the next positive example in the session, $p_{m}$, such that:

\begin{center}
$m = \min_{j}\{j \in I_{P} \mid j > i\}$ 
\end{center}
Where the difference $m - i$ represents the number of skipped tracks between track $t_{i}$ and its next positive sample. \\
\indent To predict the desired next track at time step $i$, we model a probability distribution \textbf{$p(h_{i} = t \mid S_{i})$} over all possible tracks. Sorting this distribution provides a ranking of the most-relevant items. By learning from negative feedback, we aim to both raise the ranking of $p_{m}$, as well as lower the rankings of items in $N$ in predicting each $h_{i}$.
\subsection{Model Architecture}
\textbf{}
We investigate unidirectional and bidirectional transformer-based architectures in this study, inspired by the SASRec \cite{DBLP:conf/icdm/KangM18} and BERT4Rec \cite{DBLP:conf/cikm/SunLWPLOJ19} architectures, respectively. For both approaches we use the same base architecture described below, with the sole differences being the training procedure, learning objective, and the use of a causal attention mask in the case of the unidirectional model. We keep the implementation analagous to that of the aformentioned authors for better comparison.

\subsubsection{Track Embeddings}

We store learned track embeddings in a lookup-table $e_t \in E$ of size $T \times \mathbb{R}^{|d|}$, where $T$ is the number of tracks and d is the embedding dimensionality. $E(\cdot)$ denotes the function retrieving the embeddings of a track or set of tracks from table $E$.  

\subsubsection{Positional Embeddings} To inject information about the position of each track in the sequence, we add a learnable positional embedding $PE$ of size $K \times \mathbb{R}^{|d|}$ to each track embedding in the sequence, where K corresponds to the size of the sequence.

\subsubsection{Encoder}
We employ a standard transformer encoder to learn contextual session-level information. This is a fully attention based model employing multiple multi-head self-attention layers and position-wise feedforward layers to learn contextual information from sequential inputs. 

\subsubsection{Prediction Layer}
After obtaining hidden vectors from the encoder with contextual information, we project them through a fully connected layer with GELU activation \cite{DBLP:journals/corr/HendrycksG16} to obtain predicted embeddings $\hat{y_i}$ for each $t_i \in S$ . We then compute an inner product with the embedding table and apply a sampled softmax to get a probability distribution over each track. 
\subsubsection{Sampled Softmax}
Additionally for training stability with such a large amount of classes ($\sim$ 1M tracks in this study), we employ a sampled softmax function during training. For each mini-batch for each session, we uniformly sample 1000 unseen tracks and rank the target tracks alongside these. These 1000 tracks are re-sampled each epoch, such that as training continues, the model continually learns to "rank" the target items with an increasing subset of the total tracks, as the number of unique tracks sampled for comparison increase.

\subsection{Sequential Recommendation Task}
For both approaches, we employ the same learning objective, the negative log likelihood (NLL), for training; however they differ in how this learning objective is used. 
\subsubsection{Unidirectional}
We employ the \textit{next-item prediction} task for this approach. For each $t_i \in S$, we task the model with predicting the next item in the sequence, $t_{i+1}$. We then compute log-probabilities and pass this to the NLL Loss. Additionally, attention maps are computed using a causal mask, preserving the auto-regressive nature of unidirectional transformers.
\subsubsection{Bidirectional}
We employ the \textit{cloze}, or \textit{masked language modelling  (MLM)} task for this approach. We randomly mask a proportion $p$ of each sequence with a special token [MSK] and task the model with predicting what the correct track is at these indices with a bidirectional attention map. For the sequential recommendation task, we also append the [MSK] token to the end of the sequence and set the target of this to the last track in the session targets, to ensure that this target does not appear in the bidirectional attention map.
\subsection{Skip-informed Contrastive Task}
To learn negative sequential track relationships, we employ a contrastive learning task using the skipped tracks in each listening session. We employ noise contrastive estimation with InfoNCE \cite{DBLP:journals/corr/abs-1807-03748} shown below: \\
\begin{center}
    $\mathcal{L}_{NCE}=-\mathbb{E}_X\left[\log \frac{f_k\left(\textbf{p}, \textbf{c}\right)}{\sum_{x_j \in X} f_k\left(x_{j}, \textbf{c}\right)}\right]$
\end{center}

Given a context vector $c$, positive anchor $p$ and set of noise samples $x \in X$, this loss term uses a categorical cross entropy to classify the positive anchor from the set of noise samples, given scoring function $f_k(\textbf{\textbf{x}, \textbf{c}})$.

For each track $t_i \in S$, we adapt this to our task of promoting the next true positive sample $p_m$ and penalizing all negative samples $n_j \in N$ by defining the following:
\begin{enumerate}
\item $c = e_{t_i}$ \textbf{or} $\hat{y_{i}}$
\item $X = E(N)$
\item $p = E(p_m)$
\item $f_k(\textbf{\textbf{x}, \textbf{c}}) = \dfrac {\textbf{x} \cdot \textbf{c}} {\left\| \textbf{x}\right\| _{2}\left\| \textbf{c}\right\| _{2}}$
\end{enumerate}

This maximizes the cosine similarity between the embedding $e$ of track $t_i$ and next-positive-sample $p_m$ while minimizing the similarity between $e_{t_{i}}$ and all $e_{n} \in E(N)$. Since during prediction, logits are computed by the inner product of $\hat{y_i}$ and $E$, this directly affects the rankings of $p_m$ and all $n \in N$, by drawing $t_i$ and $p_m$ closer together in the learned embedding space, and consequently pushing $t_i$ and all $n \in N$ farther away in the embedding space. We experiment with setting the context vector $c$ as both $\hat{y_i}$ and $e_{t_i}$. Setting $c = \hat{y_i}$ includes the current session context, while setting $c = e_{t_i}$ ignores current session context and instead relies solely on the overall learned representation of the track. We explore both to examine the the extent to which immediate context and contextual history affect the learning of negative preference, respectively. 
\subsection{Dataset}
For this study we use the Music Streaming Sessions Dataset (MSSD) \cite{DBLP:conf/www/BrostMJ19} for training and evaluation, which contains ~160 Million user sessions of 10 to 20 consecutively listened songs (<60 seconds between listens). These listening sessions are uniformly sampled from a variety of contexts, such as the user’s personally curated collections,
expertly curated playlists, contextual non-personalized recommendations, and personalized recommendations. \\
\indent Notably, this dataset is pseudonymized, meaning all included sessions lack a user label. Consequently, we treat each session as a new user, ignoring long term history. \\
\indent Skip labels are provided for each track in each session with strength 1-3, defined per the authors as the track "played very briefly", "played briefly", and "played mostly (but not completely)", respectively. For this study, we are primarily interested in strong negative interactions and therefore only consider tracks with skip strength 1 and 2 as negative examples in each session.\\ 
\indent Due to time and computational restraints, we uniformly sample $\sim$450K discrete sessions containing $\sim$2 million item interactions with $\sim$1 million total unique tracks to train and evaluate our models. We note that our subset of sessions contains roughly 15\% skipped tracks.

\subsection{Training Procedure}
 As with other contrastive recommendation systems \cite{unknown, 10.1145/3447548.3467102}, we simply aggregate the sequential task loss and the contrastive shown below within one single training pass
\begin{center}
$\mathcal{L} = \alpha\mathcal{L}_{NCE} + \beta\mathcal{L}_{NLL}$
\end{center}
where $\alpha$ and $\beta$ are scalar terms. We empirically tune these parameters through the validation set.
\begin{center}
\begin{table*}
\centering
 \begin{tabular}{l|ccc|ccc}
    & \textbf{{Unidir S-}} & \textbf{{Unidir S+ $(c = \hat{y_i})$}} & \textbf{{Unidir S+ $(c = e_{t_i})$}} & \textbf{{Bidir S-}} & \textbf{Bidir S+ $(c = \hat{y_i})$} & \textbf{Bidir S+ $(c = e_{t_i})$}\hfill\\
  \hline
  \textbf{HR@ 1} (\% increase)  & .2821 & .3049 (8.08\%) &  \textbf{.3073 (8.93\%)} & .1945 & .2101 (8.02\%) &  \textbf{.2339 (20.26\%)}\\
  \textbf{HR@ 5} (\% increase) & .4803 & .5042 (4.98\%)& \textbf{.5088 (5.93\%)} & .4203 &  \textbf{.4370 (3.97\%)} &  .4358 (3.69\%)\\
  \textbf{HR@ 10} (\% increase) & .5593 & .5836 (4.34\%) & \textbf{.5867 (4.90\%)} & .5135 &  \textbf{.5270 (2.63\%)} &  .5165 (0.58\%)\\
  \textbf{HR@ 20 } (\% increase) & .6316 & .6550 (3.70\%) &  \textbf{.6554 (3.77\%)} & .5989 &  \textbf{.6101 (1.87\%)} & .5911 (-1.30\%)\\
  \hline
 \end{tabular}
 \caption{Hit Ratio @ [1, 5, 10, 20] for the unidirectional and bidirectional approaches. Baselines without skip-informed contrastive learning (\textbf{S-}) are compared to the proposed contrastive learning settings (\textbf{S+}, using either session context $\hat{y_i}$ or track representation vector $e_{t_i}$ as context vector $c$). Numbers in parentheses show the relative increase in percentage of the approach over the respective baseline; bold entries mark the best performing approach within unidirectional and bidirectional architectures.\hfill}
 \label{tab:datapoints}
\end{table*}
\end{center}
\subsection{Hyperparameters and Implementation}
As our data contains variable length sessions between 10 and 20 interactions, we pad all sessions to length 20. We stack 2 encoder blocks with 8 attention heads. The embedding and hidden dimensions are both set to 128. Masking for the bidirectional model is applied per batch with proportion $p = 0.2$.  
We initialize all parameters via truncated normal sampling with $\mu = 0, \sigma = 1$ in range $[-0.02, 0.02]$.
We tune the optimal $\alpha, \beta \in [.25, .5, .75, 1]$ using the validation set and select $\alpha = 0.5, \beta = 0.5$. We use the ADAM optimizer \cite{DBLP:journals/corr/KingmaB14} with a learning rate of 0.005, selected after tuning through the validation set with $lr \in [0.0001, 0.0005, 0.001, 0.005]$. All models were  implemented in python using pytorch-lightning and trained using an NVIDIA RTX 2070 GPU.

\section{Results and Discussion}
\subsection{Evaluation}
We employ the next-item recommendation task used by \cite{DBLP:conf/icdm/KangM18, DBLP:conf/cikm/SunLWPLOJ19} for our evaluation. For each sequence, we leave out the final and penultimate items as the testing and validation targets, respectively and reserve the rest of the sequence for training. For each target, we uniformly sample 1000 unobserved tracks, where the task becomes to rank the target among these tracks. We employ the Hit Rate@K (equivalent to recall) as our evaluation metric, with $k \in [1, 5, 10, 20]$. The results are shown in Table~\ref{tab:datapoints}.


\subsection{Discussion}
We note a number of observations from our experiments. Namely: 
\begin{enumerate}
    \item The skip-informed contrastive task consistently outperforms the feedback-agnostic models, indicating that learning from negative feedback is beneficial for sequential music recommendation
    \item The unidirectional models consistently outperform the bidirectional models, with a waning performance gap as the top-K for the hit rate increases. 
    \item Using the final hidden state $\hat{y_i}$ with immediate contextual information as the context vector for the contrastive task performs similarly but consistently slightly worse than using the item embeddings. 
\end{enumerate}
Overall, we observe that our contrastive task reliably increases the hit rate in a next-item recommendation scenario, with the exception of the HR@20 for the bidirectional model using only track embeddings. 
Interestingly, even though we create a mismatch between the targets for the sequential recommendation task and contrastive task, the hit rate for the sequential recommendation task increases, inferring that optimizing for the next positive example ($p_{m}$) and next track ($t_{i + 1})$ in tandem raises the performance in selecting the next track during inference. 

\indent We also observe waning performance gains as the number of tracks in the ranking window increases, likely due to the fact that the contrastive task only relates observed tracks with each other. As the amount of unobserved tracks in the comparison increases (i.e HR@1 to HR@20), the weaker the effect of the contrastive task. Our experiments imply that the effect of learning from negative feedback in this fashion mostly affect the top ranked recommendations. \\
\indent The relatively weak performance of the BERT-like architecture may be due to the relative high density of our dataset and our relatively short sequence lengths, so training in an autoregressive manner with each sample in the training sequence per each epoch may be better for learning latent sequential track relationships. More work is likely needed to find an optimal setup using bidirectional attention with the MLM task. \\
\indent The slight performance improvement when using the track embeddings as the contextual vector for the contrastive task may imply that while immediate session-level contextual information is useful in learning from negative feedback, reducing this emphasis may provide a slightly stronger signal for preference of a user's next desired track.

\section{Conclusion and Future Work}
Overall, we have presented both a study on the use of transformer-based architectures for sequential music recommendation, as well as a contrastive-based task to learn from negative feedback. We show through our experiments that the contrastive task results in greater hit rate on both unidirectional and bidirectional architectures. 
Multiple avenues for future work arise, namely the inclusion of long-term user profiles for better modelling of long term and changing user-taste. Additionally, contextual and content information can be injected into the embeddings to learn more powerful contextual representations. An analysis of the performance on different session types and streaming behaviors (playlist, auto-generated, user-curated, etc.) \cite{10.1145/3459637.3482123} would also provide better insight into the performance in different listening contexts.

\begin{acks}
This research was funded in whole, or in part, by the Austrian Science Fund (FWF) [P33526]. For the purpose of open access, the author has applied a CC BY public copyright license to any Author Accepted Manuscript version arising from this submission.
\end{acks}
\bibliographystyle{ACM-Reference-Format}
\bibliography{ref}


\end{document}